\documentclass[sn-apa]{sn-jnl}

\jyear{2021}%

\theoremstyle{thmstyleone}%
\theoremstyle{thmstyletwo}%

\theoremstyle{thmstylethree}%

\DeclareMathOperator*{\argmin}{argmin}

\begin{document}

\title[Temporal and probabilistic comparisons of epidemic interventions]{Temporal and probabilistic comparisons of epidemic interventions}


\author*[1,2]{\fnm{Mariah C.} \sur{Boudreau}}\email{Mariah.Boudreau@uvm.edu}

\author[1]{\fnm{Andrea J.} \sur{Allen}}

\author[1]{\fnm{Nicholas J.} \sur{Roberts}}

\author[3,4,1]{\fnm{Antoine} \sur{Allard}}

\author[1,2,3,5]{\fnm{Laurent} \sur{H\'{e}bert-Dufresne}}

\affil*[1]{\orgdiv{Vermont Complex Systems Center}, \orgname{University of Vermont}, \orgaddress{\city{Burlington}, \state{VT}, \country{USA}}}

\affil[2]{\orgdiv{Department of Mathematics \& Statistics}, \orgname{University of Vermont}, \orgaddress{\city{Burlington}, \state{VT}, \country{USA}}}

\affil[3]{\orgdiv{D\'{e}partment de physique, de g\'{e}nie physique et d'optique}, \orgname{Universit\'{e} Laval}, \orgaddress{ \city{Qu\'{e}bec}, \state{Qu\'{e}bec}, \country{Canada G1V 0A6}}}

\affil[4]{\orgdiv{Centre interdisciplinaire en mod\'{e}lisation math\'{e}matique}, \orgname{Universit\'{e} Laval}, \orgaddress{ \city{Qu\'{e}bec}, \state{Qu\'{e}bec}, \country{Canada G1V 0A6}}}

\affil[5]{\orgdiv{Department of Computer Science}, \orgname{University of Vermont}, \orgaddress{\city{Burlington}, \state{VT}, \country{USA}}}


\abstract{Forecasting disease spread is a critical tool to help public health officials design and plan public health interventions.
However, the expected future state of an epidemic is not necessarily well defined as disease spread is inherently stochastic, contact patterns within a population are heterogeneous, and behaviors change. 
In this work, we use time-dependent probability generating functions (PGFs) to capture these characteristics by modeling a stochastic branching process of the spread of a disease over a network of contacts in which public health interventions are introduced over time. 
To achieve this, we define a general transmissibility equation to account for varying transmission rates (e.g. masking), recovery rates (e.g. treatment), contact patterns (e.g. social distancing) and percentage of the population immunized (e.g. vaccination). 
The resulting framework allows for a temporal and probabilistic analysis of an intervention's impact on disease spread, which match continuous-time stochastic simulations that are much more computationally expensive.
To aid policy making, we then define several metrics over which temporal and probabilistic intervention forecasts can be compared: Looking at the expected number of cases and the worst-case scenario over time, as well as the probability of reaching a critical level of cases and of not seeing any improvement following an intervention.
Given that epidemics do not always follow their average expected trajectories and that the underlying dynamics can change over time, our work paves the way for more detailed short-term forecasts of disease spread and more informed comparison of intervention strategies.}




\maketitle

\section{Introduction}

Monitoring the spread of COVID-19 is at the forefront of public health agendas as new variants emerge. 
Transmission across the globe has forced countries to mitigate the spread with their own combination of masking and social distancing \cite{chu2020physical}, restrictions on mobility \cite{aleta2020data, althouse2020unintended}, improved ventilation \cite{sun2020efficacy}, contact tracing \cite{kojaku2021effectiveness} and other local interventions. 
Even in neighboring regions, the diversity of interventions reflect differences in local policy, culture, differences in local forecasts, as well as different goals for interventions \cite{white2020state}. For example, some populations may attempt to minimize the expected number of COVID-19 transmissions while other may only wish to minimize the probability of overwhelming their healthcare system.
Whether or not these different objectives 
would lead to the same policies is unclear given the underlying randomness and uncertainty inherent to epidemic forecasting.

There are two important issues to consider when comparing forecasts of epidemic interventions: Forecasts should be \textit{probabilistic} and \textit{time-dependent} as disease spread is stochastic and heterogeneous \cite{noel2009time, allen2022predicting}. Temporal probabilistic forecasts must then be summarized by specifying given statistics, as well as a temporal window to target, chosen to capture the intended goal(s) of the intervention. And, since forecasts evolve, the relative effectiveness of two policies can itself vary over time. Altogether, comparing multiple intervention policies is not as simple as comparing the averaged effective growth rate of the epidemic.

Past work on intervention comparisons has studied how different policies such as lockdown strategies or physical distancing impact disease trajectory within a population \cite{Milne2008,Wessel2011,Peak2017,Davies2020,Churches2020}. 
Most of the comparisons in the literature, however, are based around the average of the stochastic (often simulated) outcomes or present confidences intervals for derived measures such number of hospitalizations or the effective reproductive number \cite{Milne2008,Davies2020}. 
In comparison, our philosophy is more similar to probabilistic forecasting in meteorology, where a cone of uncertainty of storm paths or expected rainfall are the targets. 
We argue that new summary statistics, which directly compare disease outcomes and their probability of occurring, need to be developed to account for the stochastic nature of disease trajectories.

In this paper, we use a mathematical framework to track the distribution of cumulative and active cases in a networked population over the course of epidemic generations. When compared to simulations, these epidemic generations offer a surprisingly accurate proxy for the actual temporal dynamics of the epidemic \cite{allen2022predicting}.  We extend this framework in Sec.~\ref{sec:methods} to allow temporal interventions that affect parameter values or contact structure from one epidemic generation to the next, thereby modifying the probabilistic epidemic forecasts over time. In Sec.\ \ref{sec:interventions}, we present specific network interventions and offer a series of summary statistics chosen to capture the different possible goals of these interventions.

\begin{figure}[t]
    \centering
    \includegraphics[width=0.6\linewidth]{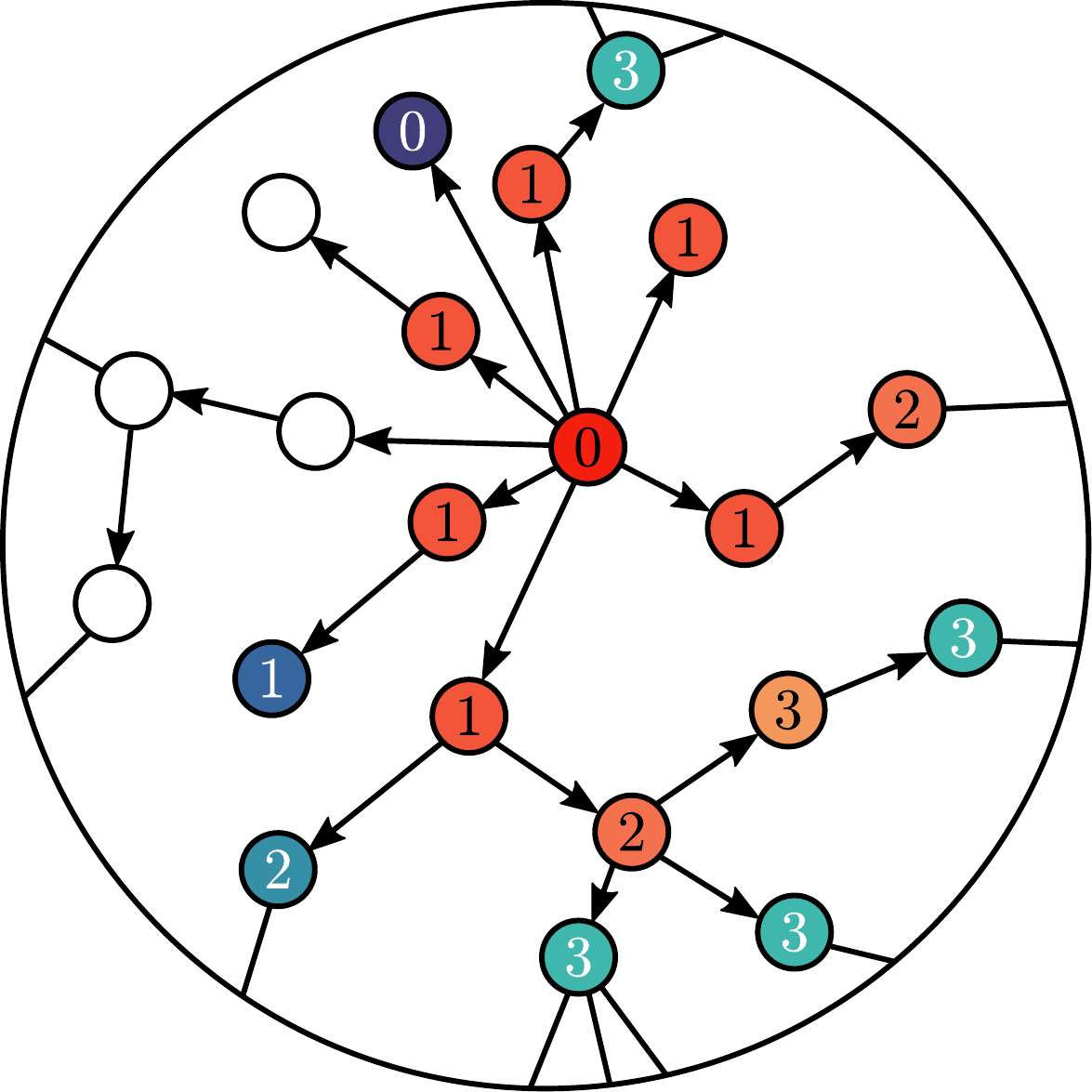}
    \caption{\textbf{Schematic of generations of infection through a network with interventions.} An initial node is infected during generation 0 (shown in deep red). Subsequent epidemic generations are represented in shades of red, with each node labeled in black by the generation in which it was infected. The blue shaded nodes were part of an intervention (e.g., vaccination), hindering the spread of the infection along that branch of the tree if the intervention preceded a potential transmission. Interventions are also temporal, shown in shades of blue and labeled in white by the epidemic generation when their intervention occurred. The branching dynamics of the resulting transmission tree are highly complex as the two dynamical processes compete, with the disease potentially spreading exponentially but slowing down as the intervention ramps up.}
    \label{fig:cartoon_network}
\end{figure}

We demonstrate our approach to a specific case study in Sec.~\ref{sec:rand-target-comparison} where we compare targeted and random vaccination rollouts. Targeted vaccination is meant to immunize highly connected individuals (e.g. healthcare workers) that are at higher risk of receiving and passing the epidemic. However, this strategy comes at a cost and we assume that the targeted rollout of a vaccine must be slower than the random rollout of the same vaccine. Using our mathematical model and our summary statistics of temporal probabilistic forecasts we then ask: How fast must targeted vaccination be to outperform random vaccination? Do different metrics of intervention performance lead to the same answer? There are complex competition dynamics occurring between the epidemics unfolding on a contact network and interventions rolled out to affect this network (see Fig.~\ref{fig:cartoon_network}). This work establishes a framework to study this dynamics and answer the previous questions. Section~\ref{sec:discussion} outlines the generality of our approach, showcasing other types of interventions which can be modeled using our methodology.

\section{Theoretical Analysis \label{sec:methods}}

\begin{figure}
    \centering
    \includegraphics[scale = 0.4]{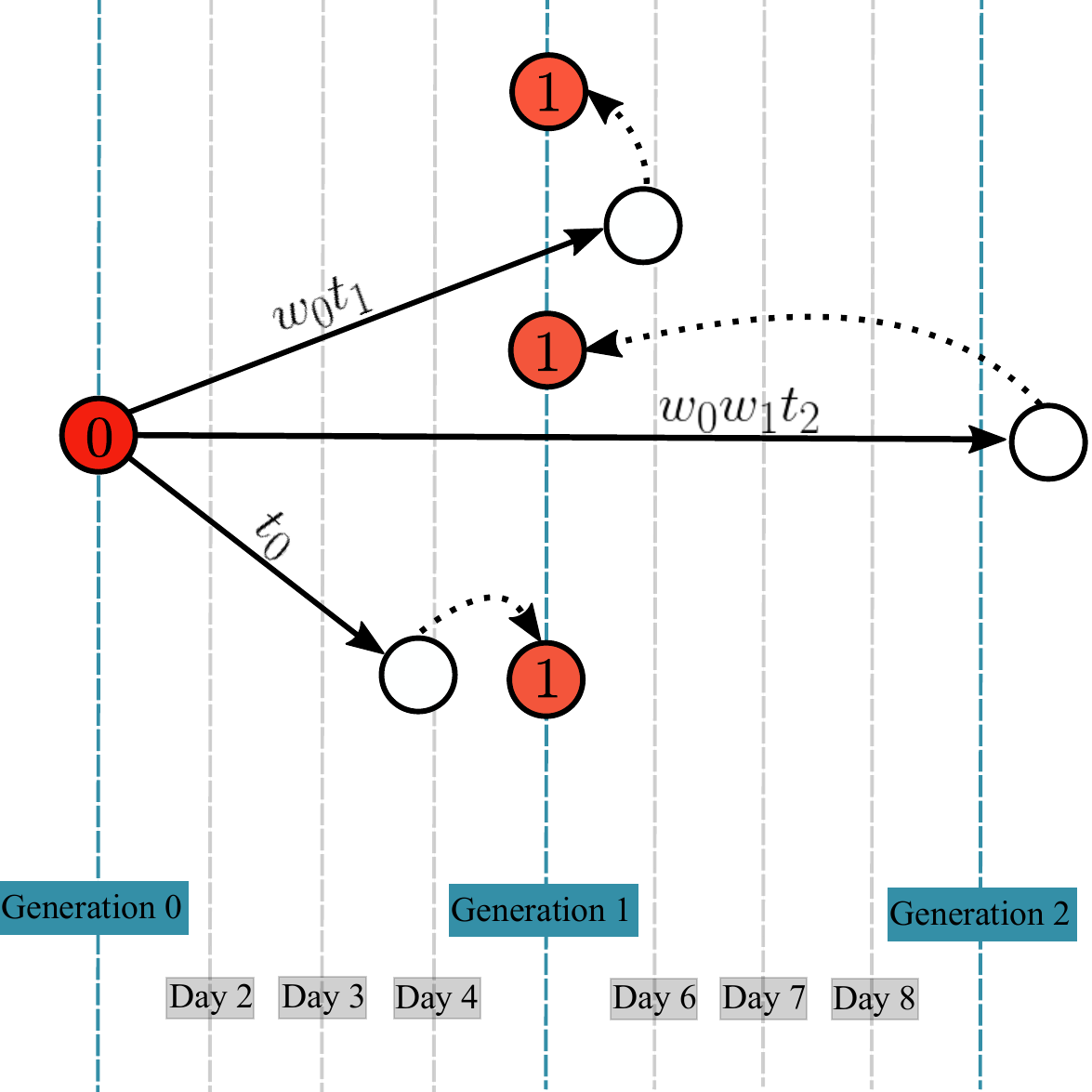}
    \caption{\textbf{Mapping continuous-time dynamics to branching process generations.} The process in which continuous-time disease spread is mapped to a discrete-time branching process is shown above. An infectious individual will infect a certain number of other individuals via a branching process, which is captured by the various transmission terms in Eq. (\ref{eq:transmission-expansion}). Once those individuals are identified, they are mapped to the next epidemic generation. For this specific example, we have an initial infectious individual (red node marked by generation 0), that infects three individuals at different probabilities of infection. If the transmission occurs in the same generational-time interval, here in the 0-th interval with probability $t_{0}$, the new case (bottom red node) becomes infectious at generation 1. When the transmission occurs during generation 1, the individual is conceptually mapped back to the start of generation 1 (top red node) and this occurs with probability $w_{0}t_{1}$. 
    This probability is the probability of the 0-th generation passing multiplied by the probability of transmission occurring during the first generational interval. 
    Likewise, there is a probability of two generations passing before a transmission occurs, with probability $w_{0}w_{1}t_{2}$, meaning the individual (middle red node) also mapped back to the start of generation 1.
    This mapping allows the analysis of continuous-time epidemic dynamics as a simpler discrete branching process.}
    \label{fig:branch-to-gen}
\end{figure}

\subsection{Assumptions}
Our framework assumes that the spreading process of the disease being studied follows undirected percolation dynamics over a contact network and can therefore be analyzed as a branching process.Even though the underlying transmission dynamics occur in continuous time, we determine the probability of infection according to discretized generational time. This is represented in Fig. ~\ref{fig:branch-to-gen}, where each solid-line arrow is a transmission event labeled with the probability of infection. We map these stochastic transmissions to a discretized epidemic generation. This discretization is shown in Fig. ~\ref{fig:branch-to-gen} with the vertical dotted lines representing time passing. At each generation, the branching process of transmissions from each infectious individual provides the new infectious individuals for the next generation. Even if transmission occurs in continuous time, the discrete-time mapping places individuals in the subsequent generation (sometimes underestimating time to transmission, sometimes overestimating it). The system is then updated and this process continues for the time-frame set. This assumption that transmission aligns with generational time makes analytical calculations and tracking of active cases easier even though it introduces small errors given that transmissions are pulled backward and forward in time, see Fig. ~\ref{fig:branch-to-gen} caption for an example case.
This is an approximation of a spreading process \cite{kenah2007second} but was recently shown to provide accurate temporal forecasts when compared to continuous-time simulations \cite{allen2022predicting}.

In our case studies, we also assume that contacts in the population follows a geometric distribution. The aim of this assumption is to have a heterogeneous network of contacts. The geometric distribution is the discrete equivalent of the exponential distribution, which has been observed in real-world contact patterns [1-2]. With this distribution, we calculate the average number of secondary cases, $R_{0}$, to be 3.
Though contact networks are inherently temporal, we here assume a static contact network except for the removal of connections due to network-based interventions. When applying an intervention, specifically a vaccination strategy, we assume that vaccination offers perfect protection.
Likewise, the intricacies of vaccine efficacy (e.g. waning of immunity or the need for multiple doses) will not be covered in this work but could be incorporated in the framework.
Our goal is instead to provide a general model of disease spread and showcase how a few specific types of interventions can be included in temporal, probabilistic, and analytical forecasts.  The software associated with our model is available at Refs. \cite{andrea_allen_2021_5076514, andrea_allen_pgf_formalism}.

Forecasts, in our framework, are defined as the time evolution of our branching process approach and will not be directly validated with data. While the final states predicted by our general approach have been previously validated with empirical data [1], data to produce temporal forecasts of interventions are not available. Further validation would require contact distributions, epidemiological parameters, and incidence rates within communities before and after interventions. Instead, we rely on simulations for validation.\\

\subsection{No\"el et al. probability generating function (PGFs) formalism\label{sec:pgfFormalism}}
PGFs allow us to include inherent heterogeneity in epidemiological forecasting by calculating the probability distribution associated with specific network transmission trees. 
Generating functions offer elegant derivations of many statistical properties \cite{wilf2005generatingfunctionology, newman2002spread}.

For epidemiological forecasting purposes, the focus is on the PGF of the network degree distribution, defined as
\begin{equation}\label{eq:g0}
    G_{0}(x) = \sum^{\infty}_{k=0} p_{k}x^{k}, 
\end{equation}
where the $k$th coefficient, $p_{k}$, is the probability of randomly choosing a node with degree $k$ from the network. The average degree of the network, $\langle k \rangle$, is found by differentiating Eq.~\eqref{eq:g0} and evaluating at $x=1$,
\begin{equation}
     G'_{0}(1) = \langle k \rangle = \sum^{\infty}_{k=0} kp_{k}.
\end{equation}
This result is used to generate the distribution of potential transmissions, or the excess degree distribution,

\begin{equation}\label{eq:G1}
     G_{1}(x)
       = \frac{G_0^\prime(x)}{G_0^\prime(1)}
       = \frac{\sum_{k}(k+1)p_{k+1}x^{k}}{\langle k \rangle}
       = \sum^{\infty}_{k=0} q_{k}x^{k}. 
\end{equation}
The probability of reaching a node with degree $k$ from a randomly chosen edge is represented by the coefficients $q_{k} \propto (k+1)p_{k+1}$ due to the fact that a node of degree $k+1$ is $k+1$ times more likely to be connected to a random edge than a node of degree 1. The node of degree $k+1$ then has $k$ remaining edges to transmit through, which corresponds to the derivative and renormalization of the original PGF.

To incorporate the disease spread through the excess degree distribution, $q_{k}$, it is necessary to include, $p_{l\vert k}$, the probability of $\ell$ transmissions from a single infectious node, given that it has excess degree $k$,
\begin{equation}\label{eq:prob_of_transmission}
    p_{l\vert k} = \binom{k}{l}T^{l}(1-T)^{k-l},
\end{equation}

where $T$ is the probability of transmission and is further explained in Sec.\ \ref{sec:manatee}. Therefore, the number of infections caused by ``patient zero" is equal to the probability of having degree $k$ and transmitting the disease to $\ell$ of those $k$ neighbors. This is defined as $G_0(x;T)$, given by
\begin{align}
    G_0(x;T) & = \sum_{l = 0}^{\infty}\sum_{k=l}^{\infty} p_{k}p_{l\vert k}x^{l} \nonumber \\
      & = \sum_{k = 0}^{\infty}\sum_{l=0}^{k} p_{k}\binom{k}{l}T^{l}(1-T)^{k - l}x^{l} \label{eq:offspring_dist} \nonumber \\
      & = G_{0}\big(Tx + (1-T)\big).
\end{align}

As $G_{1}(x)$ is derived from $G_{0}(x)$, so can $ G_{1}(x;T)$ be derived from $G_0(x;T)$. In a static network, $G_{1}(x;T)$ represents the PGF for the probability distribution of the number of infections caused by a single node, i.e., the secondary case distribution.

Now, PGFs traditionally do not keep track of time as the branching process 
unfolds; however, No\"el \textit{et al.} developed a piece-wise generating function that tracks the branching process via generations \cite{noel2009time}. Mathematically, for a static network, this is given by

\begin{equation}\label{eq:generations}
    G_{g}(x; T) = 
    \begin{cases}
    G_{0}(x; T) & g = 0\\
    G_{1}(x; T) & g > 0,\\
    \end{cases}
\end{equation}
where $G_{0}(x;T)$ defines the distribution for the first generation and $G_{1}(x;T)$ defines all future generations. In this work we expand on the framework laid out above to demonstrate the effect that temporal behaviors can have on the branching process.

Following No\"el \emph{et al.}~\cite{noel2009time}, we calculate
the cumulative case distribution. To do so, we use a simple generation scheme illustrated in Fig.~\ref{fig:cartoon_network}: Any transmission from a node infected in generation $g$ is considered to be in epidemic generation $g+1$ regardless of the exact timing of the transmission event. From this, let $s_g$ be the number of cumulative cases at generation $g$ and let $m_g$ be the number of infectious nodes strictly belonging to generation $g$. 
Note that in this way, $s_g = \sum_{g'=0}^g m_{g'}$.
We denote $\psi_{sm}^{g}$ the probability of having $s$ total infections by the end of the $g$-th generation with $m$ becoming infected (and thus being infectious) during that generation. We also denote
\begin{align}
  \Psi_0^g (x,y) = \sum_{s=1}^{\infty} \sum_{m=0}^{s} \psi_{sm}^{g}x^{s}y^{m}
\end{align}
the associated PGF. As demonstrated in Ref.~\cite{noel2009time}, $\Psi_0^g (x,y)$ 
is derived via a recursive function for the probability of $s_{g-1}$ total infections in generation $g-1$. Each new infection, $m_{g-1}$, in $g-1$ has its own possible transmission connections, $G_{g-1}(xy;T)$, incorporating all possible transmission events leading up to generation $g$. Mathematically, this is given by
\begin{align}\label{eq:totalInfectionsInG}
    \Psi_0^g (x,y) 
      & = \sum_{s'=1}^{\infty} \sum_{m'=0}^{s'} \psi_{s'm'}^{g-1}x^{s'} [G_{g-1}(xy;T)]^{m'} \nonumber \\
      & = \Psi^{g-1}_0 (x, G_{g-1}(xy;T)).
\end{align}
Successive iterations of Eq.~\eqref{eq:totalInfectionsInG} from an initial condition (e.g., $\Psi_0^0 (x,y)= xy$ for a single patient zero) then allows to compute $\psi_{sm}^{g}$ at the desired generation $g$.

\subsection{Formalism extension: altering transmission\label{sec:manatee}}

Given a time point, intervention strategies can be implemented, altering the future dynamics of the disease spread. From Eq.\ (\ref{eq:generations}), we generalize the piece-wise generating function to adhere to the intervention strategy being utilized. Given the type of intervention strategy, there can be multiple generations with an intervention implemented. So, to capture interacting temporal features of the disease spread and intervention, each epidemic generation is defined by its own PGF, 
\begin{equation}\label{eq:generationWithIntervention}
      G_g(x;T_g),
\end{equation}
as contact patterns change along with a new transmissibility expression, $T_g$, which will be derived in the following section.
We represent this model in Fig. \ref{fig:cartoon_network}, where the branching process is dynamically slowed by an intervention rollout. 

PGFs model a stochastic process which encapsulates the random nature of disease spread. The probability of a current infectious person causing a new infection, or the probability of transmission, is captured in $T$. We will follow Susceptible-Infectious-Recovered (SIR) dynamics which could depend on the time since infection $t’$, the time-dependent transmission rate $\beta(t’)$, and time-dependent recovery rate $\gamma(t’)$. One could then calculate a general probability for transmitting before recovery, but the exact calculation is often model-dependent. We will follow most models and consider that transmission and recovery as simple Poisson processes occurring at fixed rate $\beta$ and $\gamma$ respectively. However, transmission occurs only if the contact is not immune, which is true with probability $(1-V_{g})$, where $V_{g}$ is proportion of the population that has been vaccinated by generation $g$. In other words, $V_g$ is the cumulative proportion of the population vaccinated. Assuming infectiousness of a node in generation $g$ lasts for some random time \(\tau\) then the probability of the individual transmitting infection to another individual is
\begin{align}\label{eq:general_T}
    T(\tau) & = (1-V_{g})[ 1 - \lim_{\delta t \to 0}(1 - \beta \delta t)^{\tau/\delta t}] \nonumber \\ & = (1-V_{g})(1 - \exp^{- \tau \beta}) \; .
\end{align}
When evaluating the probability of a particular \(\tau\), the cumulative distribution function over $\tau$ is evaluated, shown by
\begin{align}\label{eq:cdf-derive-T}
    F(\tau) & = (1 - \lim_{\delta t \to 0}(1 - \gamma \delta t)^{\gamma/\delta t}) \nonumber \\ & = (1 - \exp^{- \gamma \tau })
\end{align}
The above derivation uses the average rate of recovery, $\gamma$. Taking the derivative of Eq. (\ref{eq:cdf-derive-T}) gives the probability mass function over \(\tau\),
\begin{equation}
    f(\tau) =\gamma \exp^{- \gamma \tau } \; .
\end{equation}
We can then compute the total probability of transmission by calculating the average probability of an individual transmitting before its recovery, given that the individual recovers at time $\tau$. The average transmissibility for a generation, $T_g$, is therefore
\begin{equation}\label{eq:general-T}
    T_g = (1-V_{g})\int_0^\infty T(\tau) f(\tau) d\tau = (1-V_{g})\frac{\beta}{\beta + \gamma},
\end{equation}
with this derivation following Refs.\ \cite{hebert2013pathogen,hebert2015complex}. The expression for $T_g$ allows us to interpret the probability of transmission as the probability a transmission occurs first in a superposition of Poisson processes, transmission and recovery with rates \(\beta\) and \(\gamma\) respectively.

In our model, the passage of time to the next generation must also be included, which is determined by the product of the average excess degree, $q = G^\prime_1(1)$, and the transmission rate yielding $q\beta$~\cite{allen2022predicting}. Allen \textit{et al.} confirms similarities in the mapping between continuous time and discretized generational time given this rate for passage of time. Section 3.3 discusses the continuous-time simulations that exhibit the validation of our generational approach. Treating this as another Poisson process, and allowing for interventions, we find the probability of a single person causing an infection leading to the next generation to be
\begin{equation}\label{eq:tg}
    t_g = (1-V_{g}) \frac{\beta}{\beta + \gamma + (1-V_{g})q_g \beta},
\end{equation}
where again $V_g$ is the cumulative proportion of the population vaccinated by at $g$, and where $q_g$ is the generation-dependent average excess degree and is defined at Eq.~\eqref{eq:updated-q}. Similarly, the probability that the next generation occurs \emph{before} a given person either transmits the disease or recovers is
\begin{equation}\label{eq:wg}
    w_g = \frac{(1-V_{g})q_{g} \beta}{\beta + \gamma + (1-V_{g})q_{g} \beta}.
\end{equation}
To encapsulate the probability transmission given Eqs.\ (\ref{eq:tg}) and (\ref{eq:wg}) for each generation, we combine the probability of a single person causing an infection leading to the next generation with the sum of probabilities that the next generation occurs before a particular transmission or recovery event,

\begin{align}
\label{eq:transmission-expansion}
	T_g &= t_g + w_{g}t_{g+1} + w_{g}w_{g+1}t_{g+2} +
	 ... \nonumber \\ 
	&= t_g + \sum_{\ell=g+1}^\infty \Big( \prod_{\ell'=g}^{\ell-1} w_{\ell'} \Big) \cdot  t_\ell.
\end{align}

This last expression closes our mapping of continuous-time SIR dynamics to a discrete-time branching process. The same recipe can be used to map other compartmental models to branching processes by including additional mechanisms. Alternatively, a SEIR with a fixed latent period of one epidemic generation can be implemented by setting $t_{g} = 0$ and $w_{g} = 1$ to delay transmission. Regardless, once transmissions dynamics are mapped to a discrete-time (or generational) branching process, our general framework is agnostic to the details of the transmission mechanisms.

\begin{figure}[t] 
    \centering
    \includegraphics[width=\linewidth]{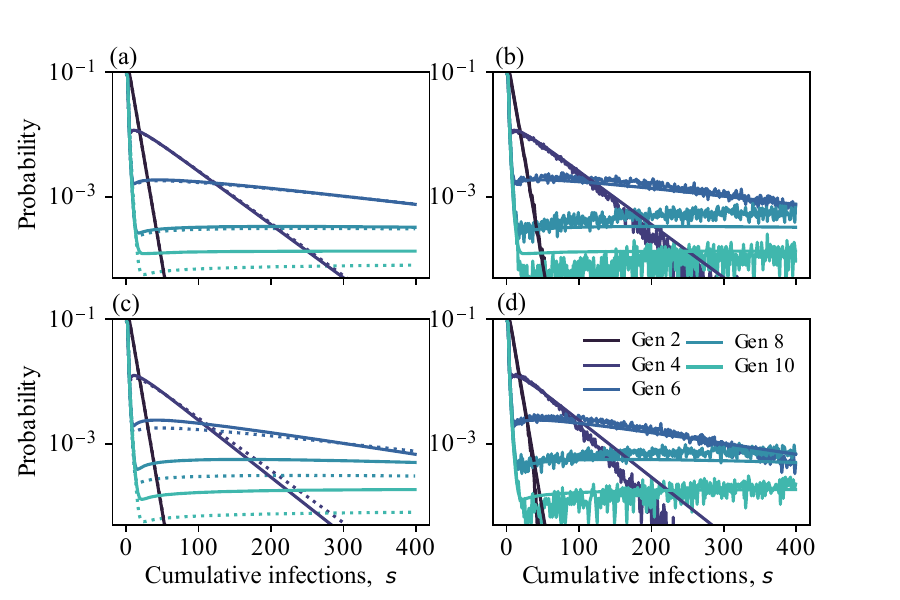}
    \caption{\textbf{Random and targeted rollout comparison and validation.} We use a geometric distribution defined by $p_{k} = 0.6^{k-1}(0.4)$, where $k = 1$, resulting in $R_{0} = 3$. Each panel details probability distributions of cumulative infections at generations 2, 4, 6, 8 and 10. \textbf{Panel (a)} depicts the comparison between a non-intervened system (dotted lines) and a random rollout strategy of 0.5\% of the population being randomly chosen to be vaccinated generations 4, 6, 8, and 10 (solid lines). By the end of generation 10, 2.0\% of the population is vaccinated. \textbf{Panel (b)} depicts simulations of the random rollout vaccination strategy, which validates the modeled generations. \textbf{Panel (c)} depicts the comparison between a non-intervened system (dotted lines) and a targeted rollout strategy where the first 0.5\% of highest degree individuals were chosen to be vaccinated at generations  4, 6, 8, and 10 (solid lines). \textbf{Panel (d)} depicts simulations of the targeted rollout vaccination strategy, which validates the modeled generations.}
    \label{fig:four-panel}
\end{figure}

\section{Interventions}\label{sec:interventions}
With an understanding of how the transmission process of this framework operates, we now aim to take individuals out of this process via intervention strategies. It is important to remember that we assume a node infected in generation $g$ infects nodes that are mapped to be in generation $g+1$. If an intervention strategy is implemented during a generation, it is assumed it would occur immediately at the start of that generation. Intervention strategies directly alter the numerical value of Eq. (\ref{eq:transmission-expansion}), then Eq. (\ref{eq:totalInfectionsInG}) is recalculated to update the probability of having $s$ cumulative infections and $m$ active cases. The variables $q_g$ and $V_g$ that appear in Eq.\ (\ref{eq:transmission-expansion}), via \(t_g\) in Eq.\ (\ref{eq:tg}) and \(w_g\) in Eq.\ (\ref{eq:wg}), correspond with one of two types of intervention strategies. Respectively, the two types of strategies are: 1) Uniform or random interventions, where proportions of the population are not susceptible to the disease. 2) Network interventions, which pertain to altering the degree distributions, such as targeted vaccination. When no network interventions are imposed on the system, $q_{g}$ is kept consistent across all generations, after it is derived from the original excess degree distribution. Conversely, the value of $V_{g}$ is kept at zero for all generations when there is a targeted intervention, since contacts around targeted vaccination are removed in $q_g$ and no contacts can then lead to vaccinated nodes in this scenario. Here we focus on uniform interventions and network interventions, along with comparing them.

\subsection{Uniform or random interventions}\label{sec:uniform-int}
In this work, we consider the uniform intervention as a \emph{random vaccination} strategy.  This intervention strategy is implemented by randomly vaccinating susceptible nodes in the population with uniform probability \cite{pastor2002immunization}. The $V_g$ term of Eq.~(\ref{eq:tg}) and Eq.~(\ref{eq:wg}) represents the probability of a node being vaccinated, along with the proportion of the population to be vaccinated at generation $g$. This quantity is therefore always a fraction between 0 and 1.

When a vaccination intervention is implemented at only one generation, meaning in a single intervention, the vaccinated population $V_g$ changes as a simple step-function. Realistically, vaccination interventions are implemented over time and over multiple generations, which we can incorporate into our modeling framework by defining a \emph{rollout} strategy. Under a vaccination rollout, the cumulative percentage of the population to be vaccinated is spread over multiple generations, slowly affecting the growth of the epidemic spread along each of its active generational branches. For multiple generations, we can state that the total proportion of the population vaccinated over all generations, or cumulative percentage vaccinated, is given by 
\begin{equation}\label{eq:cumulative-v}
    V_{total} = \sum_{g=1}^{\infty} V_g,
\end{equation}
where each generational proportion vaccinated is defined as
\begin{equation} \label{eq:general-v-g}
    V_g = \sum_{k = 0}^{\infty} \delta^{g}_{k} p_k.
\end{equation}
When implementing random vaccination, the entire network is uniformly vaccinated, resulting in

\begin{equation}\label{eq:delta-k-random}
    \delta^g_k = V_g,
\end{equation}
 for all $k$. It is important to note that random vaccination does not alter the general structure of the degree distribution or the average excess degree due to the condition set in Eq.~(\ref{eq:delta-k-random}).

Figure~\ref{fig:four-panel}(A) showcases the difference in the probability distributions of cumulative infections on a system that has no intervention implemented (dotted lines), and one with a random rollout of 0.5\% occurring at generations 4, 6, 8, and 10 (solid lines). Given that the intervention does not occur until generation 4, the distributions for generation 2 are exactly the same. For generations occurring after generation 4, the distributions begin to deviate from one another. 
 
Figure \ref{fig:four-panel}(B) validates the extended PGF formalism for multiple interventions, with the theoretical distributions shown alongside numerical simulations following an event-driven, continuous time framework~\cite{allen2022predicting}. The analytical distributions show a bit of an overestimation at generation 10, which will be discussed in Sec.~\ref{sec:validation}.

\begin{figure}[t]
    \centering
    \includegraphics[width=\linewidth]{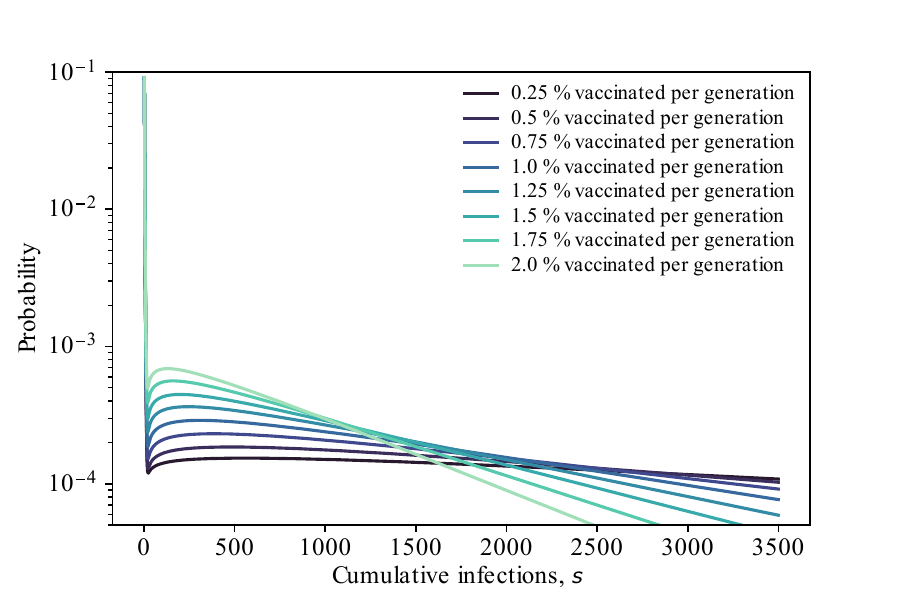}
    \caption{\textbf{Flat distributions at generation 10.} Given a geometric distribution defined by $p_{k} = 0.4^{k-1}(0.6)$, each line represents the probability distribution of cumulative infections at generation 10. The difference between the distributions is that the percentage of the population that were chosen to be vaccinated at generations  4, 6, 8, and 10 varies. The lower percentages per generation lead to flat distributions, whereas the higher percentages per generation provide distributions that have zero probability of cumulative infections past a certain point.}
    \label{fig:crossover}
\end{figure}

\subsection{Targeted network interventions}\label{sec:network-inter}
To demonstrate a network intervention, in this work we show how a \emph{targeted vaccination} strategy is implemented. The goal of targeted vaccination is to focus vaccination efforts on the group of nodes with the highest degrees in the network, or the individuals with the most contacts. This strategy results in reducing the impact of the individuals that have the most potential for creating a superspreading event, for an example see \cite{rosenblatt2020immunization}.

Given a percentage of the population to vaccinate in $g$, as defined in Eq.~(\ref{eq:general-v-g}), we start with degree classes $k' = k_{max}$ and vaccinate a fraction $\delta_{k'}$ of the degree class before moving to degree class $k'-1$. To determine the fraction vaccinated for each degree class, we define
\begin{equation}
    \delta^{g}_{k'} = \begin{cases}
        1, & \text{if } \sum^{\infty}_{k=k'} p_k < V_g \\
        \bigg(V_g - \sum^{\infty}_{k=k'+1} p_k\bigg) 
 p_{k'}, & \text{if } \sum^{\infty}_{k=k'+1} p_k < V_g \\ 
        & \text{ and } \sum^{\infty}_{k=k'} p_k > V_g\\
        0, & \text{otherwise,}
    \end{cases}
    \nonumber
\end{equation}

where this non-uniform intervention will alter the degree distribution and average excess degree.

By the independence assumption of the configuration model, each neighbor will be vaccinated in generation $g$ with probability equal to the probability that following a random edge leads to a vaccinated node, call this $H_g$. Thinking in terms of number of edges we thus compute 
\begin{equation}\label{eq:h-g}
    H_g = \frac{\sum_{k =0}^{\infty} (k+1)\delta_{k+1} p_{k+1}}{\sum_{k = 0}^{\infty} (k+1)p_{k+1}}.
\end{equation}
Therefore, the probability of a node being unvaccinated in $g$ is equal to $1-H_g$. 

Now, the truncation of the degree distribution in $g$ alters the average excess degree $q_g$, along with the coefficients of Eqs.~(\ref{eq:g0}) and (\ref{eq:G1}). To determine the new $q_g$, we must recompute $G'_{g}(1)$. Multiplying this by the proportion of nodes that are unvaccinated gives an updated $q_g$,
\begin{equation}\label{eq:updated-q}
    G'_{g}(1) = (1 - H_g)\bigg[\frac{\sum_{k = 0}^{\infty} k(k+1)(1-\delta_{k+1}) p_{k+1}}{\sum_{k = 0}^{\infty} (k+1)(1-\delta_{k+1})p_{k+1}}\bigg] = q_g,
\end{equation}
which is then used to derive the new $T_{g}$ for a given $g$.

Similar to random vaccination described in Sec.~\ref{sec:uniform-int}, targeted vaccination can be implemented via one instance of vaccination, or multiple. Figure~\ref{fig:four-panel}(C) shows the difference between a non-intervention strategy and a targeted rollout vaccination scheme of 0.5\% at generations 4, 6, 8, and 10. A rollout strategy is conducted in the same manner for both random and targeted vaccination. Similar to random vaccination, the non-intervention leads a lower probability of seeing 100-400 cases than targeted (or random) vaccination. Does that mean the non-intervention is better?  This question is answered in Fig.~\ref{fig:crossover}, which showcases how the weaker the intervention, the flatter the cumulative case distribution. These flatter distributions allow for there to be some chance of infecting more individuals over time.  Figure \ref{fig:crossover} also shows that the stronger intervention, the more probability mass accumulates towards the smaller cumulative infection counts. This explains why interventions appear to do worse than no intervention at smaller values of cumulative case count. Even though this figure utilizes a targeted rollout vaccination strategy the same relationship holds for random vaccination strategies. \\

\subsection{Validation via simulations}\label{sec:validation}

The simulations shown in Figs.~\ref{fig:four-panel}(B) and (D) used to validate the theoretical framework were performed using an event-driven, continuous time approach on 150 distinct networks of 20,000 nodes with 500 simulations run on each network. This totals to 75,000 simulations per validation.

The analytical distributions of infection under different vaccination strategies capture the relationships between generations of infection, but tend to overestimate the number of cumulative infections compared to the continuous time simulations. This is due to a few factors; primarily, the finite-size effects of simulations on networks with 20,000 nodes, which support faster computational time but 
results in a sharper decrease in the size of epidemic generations than are captured by the branching process model, as discussed in Ref.~\cite{allen2022predicting}. Nonetheless, the important behavior of the distributions are captured in relation to one another, and across different vaccination strategies, allowing for comparison and ranking of the effects between strategies regardless of slight numerical precision errors.

Another source of discrepancy between the model and continuous time simulations is the inability of the model to account precisely for already-infected nodes by the time of an intervention. This quantity is estimated in Eq.~\eqref{eq:transmission-expansion}, but may result in slight differences to the continuous-time simulations under which nodes who are already infected or recovered and are identified for targeted vaccination are not excluded, and are just ignored. This problem arises more for targeted vaccination efforts than random, since nodes in the targeted high degree classes are the same nodes that are likely to have been infected early in the spreading process. The results observed in simulation may experience a reduced disease burden on the population than the theoretical model which assumes an infinite supply of these high-degree nodes, because the simulation on a finite network has already burned through its supply of high-degree nodes, rendering them recovered before the vaccination intervention. 

\subsection{Comparison of interventions}\label{compare-interv}
A simple comparison for differing interventions is a direct comparison of their cumulative probability distributions at a specific generation. Beyond this, there are metrics derived from the cumulative probability distributions that provide valuable information for decision makers. 

\paragraph*{Average cases} First, we look at the expected number of cases over time, which we can denote as $\mathcal{X}^{g}_{mean}$. This corresponds to the typical approach using deterministic models that track the expected state of epidemics. Mathematically, this is defined by

\begin{equation}
    \mathcal{X}^g_{mean} = \sum_{s=1}^{\infty} \sum_{m=0}^{s} s\psi^{g}_{sm}
                         = \sum_{s=1}^{\infty} s\psi^{g}_{s},
\end{equation}
where $\psi^{g}_{s}$ is the probability of $s$ cumulative infections up to generation $g$. 

\paragraph*{Best - worst case} The second metric looks at the worst case scenario over time as a measure of the underlying heterogeneity of possible epidemic sizes. This allows us to quantify what is the largest epidemic that has a realistic probability of occurring (set by some threshold probability, $p_{t}$) and therefore to select policies that offer the ``best worst-case scenario'' for robust decision making \cite{lempert2010robust}. We derived this metric, denoted by $\mathcal{X}^{g}_{worst}$, by determining the corresponding cumulative case value, $s'$, that has probability $p_{t}$. Mathematically this means, 

\begin{equation}
  \mathcal{X}_{worst}^g = \argmin_{s'} \left\lbrace \bigg\vert\sum_{s = s'}^\infty \psi^{g}_{s} - p_{t}\bigg\vert  \right\rbrace.
\end{equation}

\paragraph*{Critical level of cases} For the third metric, we look at the probability of being at or above a critical level of cases, $c$, defined by $\mathcal{P}^{g}_{flat}$. This metric is meant to capture the common goal of \textit{flattening the curve} to avoid overwhelming the healthcare system \cite{block2020social}. From the cumulative probability distribution for cases, we derive
\begin{equation}
    \mathcal{P}^{g}_{flat}  = \sum_{s=c}^{\infty}\psi^{g}_{s}.
\end{equation}

\paragraph*{Minimal effect} Finally, we have a metric that measures the probability that a realization of an epidemic with an intervention is actually worse than a realization of the same epidemic in the same population without the intervention, $\mathcal{P}^{g}_{worse}$. This summary statistic is meant to capture the fact that some interventions might have minimal effect on the expected spread of the disease which can easily be overshadowed by the intrinsic randomness of epidemics. Mathematically we define this as 
\begin{equation}
    \mathcal{P}_{worse}^{g} = \sum_{s=1}^{\infty}\bar{\psi}_{s}^{g}\chi^{g}_{s}, 
\end{equation}
where $\bar{\psi}_{s}^{g}$ is the probability of $s$ cumulative cases when an intervention is implemented, and where $\chi^{g}_{s} = \sum_{i=1}^s \psi_{s}^{g}$
the probability of having less than or equal to $s$ cumulative cases when there is no intervention implemented.

All of the metrics above provide varying emphasis on the information from the probability distributions for cumulative cases.

\section{Case study: Random vs targeted vaccination\label{sec:rand-target-comparison}}
Given multiple vaccination strategies, choosing the best strategy involves comparing the impacts on the epidemic spreading process given some comparison criteria. 

One could compare the random rollout strategy against targeted rollout strategy in  Figs.~\ref{fig:four-panel}(B) and (D). However, given the infinite distributions computed, we cannot see much of a difference in later generations between the two strategies unless we display more than 400 cumulative infections. This is also due to the fact that is showcased in Fig.\ \ref{fig:crossover}, where the distributions could cross over each other past 400 cumulative infections.

When deciding on the best course of action and only considering the probability distributions for cumulative case counts, our results depict similar outcomes for a random and targeted vaccination strategy. With the same percentage of the population vaccinated, targeted rollout scheme proves slightly more effective, when focusing on larger cumulative cases in generations 2, 4 and 6. If the goal is to stop the spread of a disease early on, say by generation 6, a targeted rollout with a high percentage (\(V_g\)) 
 per generation utilized at the given intervention generations needs to be implemented according to this model. This is only an example of a strategy determined by the distributions of cumulative cases. Other calculations can be performed on probability distributions of cumulative cases, which can inform decision makers on different courses of action, depending on desired goals in the beginning stages of an epidemic.

\begin{figure*}[hbtp]
    \centering
    \includegraphics[width=\linewidth]{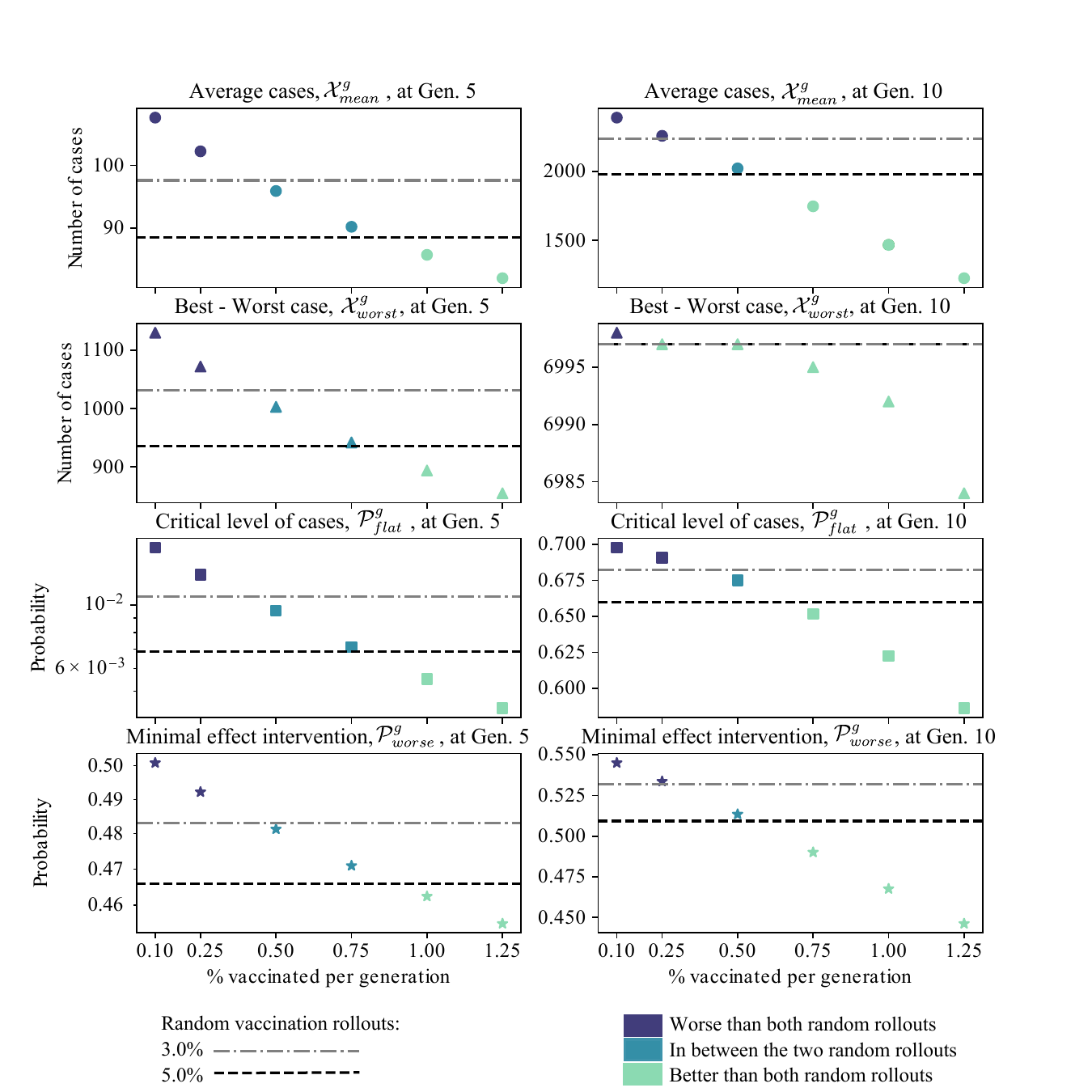}
    \caption{\textbf{Varying targeted vaccination metrics compared to two random vaccination metrics.} Given 0.10, 0.25, 0.50, 0.75, 1.00, and 1.25\% targeted rollouts per generation (rollout occurring at generations 4, 6, 8, and 10) cumulative case probability distributions, the metrics defined in Sec. \ref{sec:interventions} are computed and along with the metric for a random rollout at 3.0\% (dash-dot line) and 5.0\% (dash-dash line). The differing colors of the makers represent whether the given targeted rollout is worse than, in between, or better than the two random rollouts, as shown in the legend. The threshold for the best-worst case metric is $10^{-4}$. The critical level of cases is defined as 500 cases.}
    \label{fig:metrics}
\end{figure*}

In the previous paragraphs, a comparison of distributions provides an overall comparison, however, we are able to calculate other metrics of comparison for differing vaccination strategies. The metrics defined in Sec.\ \ref{sec:interventions} appear in Fig.\ \ref{fig:metrics}, which displays all of the metrics for targeted rollout with six different vaccination percentages ranging between 0.1 and 1.25\%. Thus there are six different cumulative vaccination strategies represented in the figure. These percentages are applied to generations 4, 6, 8, and 10, hence total proportion vaccinated is cumulative; \emph{e.g.}\ in the 0.1\% case, 0.4\% of the population will be vaccinated at generation 10. The horizontal lines represent random rollout vaccinations at 3.0\% and 5.0\%, which were rolled out at the same generations as the targeted strategy.

Figure \ref{fig:metrics} depicts a decrease in each metrics quantities as the percentage increases for a targeted rollout.  For each metric, we observe that targeted vaccination with a 0.1\% at each rollout generation is the only percentage that provides higher cases and probabilities for all the metrics of both random rollout strategies.  All of the metrics are on varying scales, yet each one follows the decreasing trend as stronger targeted rollouts are applied. The scales for between generation 5 and 10 are drastically different as well, showing how a targeted vaccination increase of 0.25\% drops the given summary statistic much more in generation 10 than generation 5. The 0.1\% and 0.25\% targeted rollout metrics are larger than the 5.0\% random rollouts for almost all the measures at generation 10, however, the difference in resources between vaccinating a total of 0.4\% or 1.0\% is much smaller than vaccinating a total of 15\% or 25\% of the population to have a similar effect. The 0.5\% and 0.75\% targeted rollouts sit in the region between the two random rollouts for all the measures at generation 5.  At generation 10, all of the metrics have the top three strongest targeted rollouts below both random vaccination strategies, along with the random vaccination lines moving closer together on their given scales. We can even see the random vaccination lines being virtually the same at generation 10 given the scale of the number of cases. These calculations allow for decision makers to evaluate what strategy will best achieve their prioritized goal during an outbreak. 

Overall, we find that there are three important factors influencing how fast a targeted roll-out must be to outperform a faster random rollout. First, the answer obviously depends on the speed of the random rollout itself. This intuitively makes sense due to the nature of targeted vaccination focusing on the individuals with the most connections. Second, it also depends on the desired temporal window: A targeted rollout of 0.75\% per generation performed worse than a 5\% random rollout per generation in all metrics at generation 5, but outperformed in all metrics by generation 10. Third, the preferred metrics of performance also influences the evaluation of interventions:  Compared to a random rollout of 5\% per generation, a targeted rollout of 0.75\% per generation minimizes all the metrics. However, one should always consider relative differences between random and targeted rollouts to fully inform decision making.

\section{Discussion\label{sec:discussion}}
%

Once a decision has been made on whether or not to implement an intervention, the question of which strategy to use arises.
Without a comparison of intervention strategies, decision makers may be lead to choose a scheme that does not mitigate the greatest concerns of their communities. Given the analytical derivations from this work, it is apparent that a targeted vaccination strategy has a faster impact on the spread of disease than a random vaccination strategy. The choice between a single instance intervention and a rollout of interventions depends on the resources of the community under consideration. Some of the vaccination strategies that only intervene in one generation may not be feasible because there is simply not enough time to vaccinate 0.15\% of the population during a generational window, let alone 3.0\% for multiple windows. This fact, along with the difficulty of determining the superspreading events in a community, makes targeted vaccination harder to coordinate than random. The relative costs of different strategies and which strategy makes the most sense in terms of resources and time is not within the scope of this paper, but are important aspects for public health officials to consider.

Although this work does not evaluate all the intricate parts of implementing various intervention strategies, it successfully captures the stochastic nature of disease spread and the heterogeneity of contact patterns and human behaviors. Due to its generational time aspect, this temporal and stochastic model removes some of the assumptions in other forecasting models, which aim to derive random disease spread, along with the impediments to the spread, over time. Another advantage to this analytical model is the transmission expression defined in Sec.\ \ref{sec:manatee} has the flexibility to accommodate intervention strategies other than uniform or network interventions. Equations (\ref{eq:tg}) and (\ref{eq:wg}) can accommodate interventions such as treatment and transmission based interventions. Values for $\gamma$ could depend on therapeutics, or a number of other types of treatments while values for $\beta$ could depend on masking, ventilation improvements, social distancing, or testing. Altogether, Eq.\ (\ref{eq:transmission-expansion}) provides a flexible approximation to account for multiple interventions, and even combinations of interventions, in probabilistic forecasts. 
Comparing interventions is a multidimensional problems, and therefore so is the design of interventions. Future work should include testing other intervention strategies, along with combining multiple strategies as we have seen happen around the world. Public health tools and forecasts need to be as heterogeneous and complex as the epidemics they aim to control.

\bmhead{Data availability statement}

All data and code associated with this project are available at Refs. \cite{andrea_allen_2021_5076514, andrea_allen_pgf_formalism}.

\bmhead{Acknowledgments}

M.C.B. is supported as a Fellow of the National Science Foundation under NRT award DGE-1735316. A.J.A. and L.H.-D. acknowledge financial support from the National Institutes of Health 1P20 GM125498-01 Centers of Biomedical Research Excellence Award and N.J.R. is supported by the University of Vermont. A.A. acknowledges financial support from the Sentinelle Nord initiative of the Canada First Research Excellence Fund and from the Natural Sciences and Engineering Research Council of Canada (project 2019-05183).

\bibliography{source}







\end{document}